\documentclass[graybox]{svmult}

\usepackage{mathptmx}       
\usepackage{helvet}         
\usepackage{courier}        
\usepackage{type1cm}        
\usepackage{makeidx}         
\usepackage{graphicx}        
\usepackage{multicol}        
\usepackage[bottom]{footmisc}

\usepackage{natbib}

\makeindex             
\begin{document}

\title*{The Orbits and Total Mass of the Magellanic Clouds}
\author{Gurtina Besla}
\institute{Gurtina Besla \at Steward Observatory, University of Arizona, 933 North Cherry Avenue, Tucson, AZ 85721, 
\email{gbesla@email.arizona.edu}}
%
%
\maketitle

\abstract{ This proceeding overviews our current understanding of the orbital history and mass of the 
Large and Small Magellanic Clouds. Specifically I will argue that the Clouds are on their first infall 
about our Milky Way and that their total masses are necessarily $\sim$10 times
larger than traditionally estimated.   This conclusion is based on the recently revised HST proper 
motions of the Clouds and arguments concerning the binary status of the LMC-SMC pair and their  
baryon fractions.  }

\section{Introduction}
\label{sec:1}

Owing to their proximity to our Galaxy, the Magellanic Clouds (MCs) have been observed in 
wavebands spanning almost the entire electromagnetic spectrum,  
allowing us to study the interstellar medium (ISM) of two
entire galaxies in unprecedented detail. Such observations have facilitated studies of 
how turbulence, stellar feedback and star formation are interrelated and how these 
internal processes affect galactic
structure on small to large scales \citep[e.g.,][]{Elm01, Block10}.

However, the MCs are also subject to environmental processes that can 
dramatically alter their internal structure.  For example, 
the MCs are surrounded by a massive complex of HI gas in the form of a 150 degree long stream trailing 
behind them (the Magellanic Stream), a gaseous bridge connecting them (the Magellanic Bridge)
and an HI complex that leads them (the Leading Arm)~\citep{Nidever10}. 
 This material
 once resided within the MCs and was likely stripped out by some combination
of external tides and/or hydrodynamic processes.

Recently, ~\citet{Fox14} revealed that these HI structures harbor a significant amount of ionized gas, increasing 
the total gas mass budget {\it outside} the MCs from $4.87 \times 10^{8}$ M$_\odot$ 
to $\sim2\times 10^{9}$ M$_\odot$.  This extended gas complex thus
 represents a non-negligible fraction of the MW's circumgalactic medium (CGM). 

Identifying the formation mechanism of these structures
depends sensitively on the amount of time the MCs have 
spent in close proximity to the MW.  Constraining the dynamics of the MCs 
is thus critical to our understanding of the morphologies, star formation histories and 
ISM properties of these important galactic laboratories.  

Our understanding of the orbital history of the MCs has evolved considerably 
over the past 10 years. The canonical view, wherein the MCs have completed multiple 
orbits about the MW over a Hubble time \citep{MF80}, has changed to one where they
are recent interlopers, just completing their first passage about our Galaxy \citep{B07}. 

This dramatic change has been driven by two factors.  
Firstly, high precision proper motions measured using the Hubble Space Telescope (HST) 
have enabled accurate 3D velocities of both the Large and Small Magellanic Clouds
(LMC and SMC). We now know the MCs are moving faster than previously believed,
relative to not only the MW, but also to {\it each other} \citep{K06a,K06b, K13}.   
 
Secondly, our understanding of the mass and structure of galactic dark matter halos has evolved.
In the $\Lambda$ Cold Dark Matter paradigm, low mass galaxies reside within
massive dark matter halos, having much larger 
mass-to-light ratios than expected for galaxies like the MW. 
This means that the MCs are likely significantly more massive than
 traditionally modeled.
Furthermore, the dark matter halos of massive galaxies are now understood to be poorly represented by
isothermal sphere profiles at large distances. Instead, the dark matter density profile falls off more
sharply, making it easier for satellites to travel to larger Galactocentric distances.    

However, debate still ensues concerning the orbital history of the MCs. While the canonical 
picture, where the MCs have completed $\sim$6 orbits about the MW with an orbital period of $\sim$2 Gyr,
has been largely dismissed, there are new proposed models where the MCs have completed one or 
two orbits about the MW within a Hubble time \citep{Sha09, Zha12, Diaz11, Diaz12}.  The goal of this review 
is to explain why the controversy arises and why various lines of evidence support a 
first infall scenario.

\section{Determining the Orbit of the MCs} 
\label{sec:3} 

Reconstructing the past orbital history of the MCs depends on 3 important factors. 1) An accurate 
measurement of the current 3D velocity vector and distance of the MCs with respect to the MW.  
2) The mass of the MW and its evolution over time. 3) The masses of the MCs, which ultimately 
determines the importance of dynamical friction as the MCs orbit about
of the MW and each other.  
\\
\newline
\noindent {\bf The 3D Velocity of the MCs:}

Recently, \citet[][ hereafter K13]{K13} used HST to measure the proper motions of stars in the LMC 
with respect to background quasars, obtaining 3 epochs of data spanning a baseline of $\sim$7 
years and proper motion random errors of only 1-2\% per field. This astonishing accuracy is 
sufficient to explore the internal stellar dynamics of the LMC, allowing for the first constraints on 
the large-scale rotation of {\it any} galaxy based on full 3D velocity measurements \citep{van14}.
This analysis has resulted in the most accurate measurement of the 3D Galactocentric velocity of 
the LMC and SMC to date. The LMC is currently moving at 321 $\pm$ 23 km/s with respect to the MW.
The SMC is moving at 217 $\pm$ 26 km/s with respect to the MW and 128 $\pm$ 32 km/s with respect
to the LMC; the SMC cannot be on a circular orbit about the LMC.   
Errors on the velocity measurement are now limited by the errors on the distance
measurement to the Clouds rather than the proper motions. 
\\
\newline
\noindent {\bf  The Mass of the MW:} 

The mass of the MW is uncertain within a factor of $\sim$2.  Values for the 
virial mass range from M$_{\rm vir}=$(0.75-2.25) $\times 10^{12}$ M$_\odot$. 
Here, M$_{\rm vir}$ is defined as the 
mass enclosed within the radius where the dark matter density is 
$\Delta_{\rm vir} =$360 times the average matter density, 
$\Omega_m \rho_{\rm crit}$.  
 
HST proper motions over a six year baseline revealed that the Leo I 
satelite is moving with a Galactocentric velocity of 196.0 $\pm$ 19.4 km/s \citep{Sohn13}.  
At 260 kpc away, this is faster than the local escape speed of $\sim$180 km/s for a 
M$_{\rm vir}$=$10^{12}$ M$_\odot$ MW model. 
Since unbound 
satellite orbits are statistically improbable within $\Lambda$CDM cosmology
\citep{BK13},  
we do not explore MW models lower than $10^{12}$ M$_\odot$ .

Few upper limits on M$_{\rm vir}$ exist apart from the timing argument, 
which limits the combined total mass of the MW and M31. 
Using the HST proper motions of M31 and
other mass arguments in the literature, \citet{van12} estimate the Local 
Group mass to be 3.17 $\pm$ 0.57 $\times 10^{12}$ M$_\odot$.  
It is thus unlikely that the MW individually contributes much more than 
$2 \times 10^{12}$ M$_\odot$. 
 
In the orbital analysis that follows, we explore 3 different mass models: 
$10^{12}$, $1.5 \times 10^{12}$ and $2 \times 10^{12}$ M$_\odot$. 
Using WMAP7 cosmology, the corresponding virial radii are 
R$_{\rm vir}$ = 250, 300 and 330 kpc.   The MW is modeled as
a static, axisymmetric, three-component model consisting of dark
matter halo, exponential disk, and spheroidal bulge.  
Model parameters are listed in Table 2 of K13. 

Note that the MW mass is expected to have grown by roughly a factor of 2 
over the past 6 Gyr \citep{Fak10}.  K13 found that this mass evolution causes the orbital 
period of the LMC to increase substantially relative to static models.
The orbital periods discussed in the following sections are thus
underestimated. 
\\
\newline
\noindent {\bf The Mass of the LMC:}

K13 found that the LMC's mass is the 
dominant uncertainty in its orbital history, since dynamical friction, 
which is proportional to the satellite mass squared, changes the 
LMC's orbit on timescales shorter than, e.g., the MW's mass evolution. 
The mass of the LMC also controls the orbit of the SMC, ultimately 
determining how long the two galaxies have interacted with each 
other as a binary pair (see $\S$\ref{sec:5}).

The LMC has a well defined rotation curve that peaks at 
Vc = 91.7 $\pm$ 18.8 km/s and remains flat out to at least 8.7 kpc 
\citep{van14}, consistent with the baryonic Tully-Fisher relation. 
This implies a minimum enclosed total mass of 
M(8.7 kpc) = 1.7 $\pm 10^{10}$ M$_\odot$; 
the LMC is dark matter dominated. 

There is strong evidence that the stellar disk of the LMC extends to
15 kpc \citep{Saha10}. If the rotation curve stays flat to at least this 
distance then the total mass enclosed is M(15 kpc) =
 $GVc^2/r  \sim 3 \times 10^{10}$ M$_\odot$. 
This minimum value is consistent with LMC masses adopted by earlier models 
\citep[e.g., ][]{GN96}.  

The total dynamical mass of the 
LMC can also be estimated using its baryon fraction.  Currently, the LMC has 
a stellar mass of $2.7\times10^{9}$ M$_\odot$ and a gas mass 
of $5.0 \times 10^{8}$ M$_\odot$. The baryonic mass 
of the LMC is thus M$_{\rm bar} = 3.2 \times 10^{9}$ M$_\odot$. 
Using the minimum total mass of  M$_{\rm tot} = 3\times10^{10}$ M$_\odot$, 
the baryon fraction of the LMC becomes M$_{\rm bar}$/M$_{\rm tot}$ = 11\%. 
This is much higher than the baryon 
fraction of disks in galaxies like the MW, which is on the order of 
3-5\%.
In the shallower halo 
potentials of dwarf galaxies, stellar winds should be more 
efficient, making baryon fractions even lower, not higher.  

This analysis is further complicated if material has been removed from the
LMC.  
As mentioned earlier, \citet{Fox14} have recently estimated the total 
gas mass (HI and ionized gas) outside the MCs at
$2\times10^{9} (d/55 {\rm kpc})^2 $ M$_\odot$. 
If half of this material came from the LMC, as suggested by \citet{Nidever08},
its initial baryon fraction would be 14\%, approaching the cosmic value.
Note that the bulk of the Magellanic Stream likely resides at distances 
of order 100 kpc, rather than 55 kpc, in which case the baryon fraction would 
increase to $\sim$20\%. 

To get a baryon fraction matching observational expectations of 
$f_{\rm bar} \sim$3-5\%, the total mass of the LMC (at least at infall) 
needs to have been $20-6 \times 10^{10}$ M$_\odot$. 
This higher total mass is consistent with cosmological expectations from 
halo occupation models that relate a galaxy's observed stellar mass 
to its halo mass.  Using relations from \citet{Mos13}, the mean halo mass for a 
galaxy with a stellar mass of $2.7 \times 10^{9}$ M$_\odot$ is 
$1.7 \times 10^{11}$ M$_\odot$, implying a baryon fraction of $f_{\rm bar}\sim$ 2-4\%
 (see Table~\ref{tab:1}).
Because there is large scatter in halo occupation models, we consider a maximal halo 
mass for the LMC of $2.5 \times 10^{11}$ M$_\odot$ in the analysis that 
follows.  
\\
\newline
\noindent {\bf The Mass of the SMC}

The current dynamical mass of the SMC within 3 kpc is constrained between
2.7-5.1 $\times 10^{9}$ M$_\odot$, using the velocity dispersion of old stars \citep{Harris06}.
This is larger than the current gas mass of the SMC is $4.2 \times 10^{8}$ M$_\odot$
and stellar mass of $3.1 \times 10^{8}$ M$_\odot$; the SMC is dark matter dominated.  
In most orbital models, the total mass of the SMC is estimated at 
$M_{\rm DM} = 1.4 - 3 \times 10^{9}$ M$_\odot$ \citep{B07, K13, Diaz11, GN96}. 
This yields $f_{\rm bar}\sim$50\%-20\%, well above cosmological expectations. 

This issue gets a lot worse when we account for the substantial amount of gas the SMC must 
have lost to form the Magellanic Stream, Bridge and Leading Arm. If we estimate again
that half the total gas mass outside the MCs comes from the SMC,
then the initial baryon mass of the SMC must have been $M_{\rm bar} = 1.73 \times10^{9}$ M$_\odot$.
Taking the traditional dark matter mass of $3\times 10^{9}$ M$_\odot$ yields $f_{\rm bar} \sim$ 60\%.  
It would take a total mass order 
$M_{\rm SMC} = 3\times 10^{10}$ M$_\odot$ to obtain a baryon fraction of 5\%.  This value is consistent 
with the choice of $M_{\rm SMC} = 2\times 10^{10}$ M$_\odot$ in B12 and B10 (see Table~\ref{tab:1}). 
Using cosmological halo occupation models the SMC dark matter mass is expected to be even higher. 
The mean expectation from relations in \citep{Mos13} is $M_{\rm SMC} = 4.2\times10^{10}$ M$_\odot$.

\begin{table}
\caption{The Total Mass of the Clouds}
\label{tab:1}
\begin{tabular}{p{2.5cm}p{2.5cm}p{2cm}p{4cm}}
\hline\noalign{\smallskip}
Baryonic Mass ($10^9$ M$_\odot$) & Dark Matter Mass ($10^{10}$ M$_\odot$)& Baryon Fraction$^a$  & Motivation for Dark Matter Mass \\
\noalign{\smallskip}\svhline\noalign{\smallskip}
{\bf LMC}  & 	& 	& 	\\
3.2  & 3 &  0.11   & Traditional Models\\
	& 6  &  0.05     &   Minimum mass to get f$_{\rm bar}<$ \%5   \\ 
        & 17 &  0.02   &    B12,  Mean $\Lambda$CDM$^c$  \\
        & 25  &  0.01  &   Maximal Model  \\ 
\hline
4.2-6.5$^b$ &  3 & 0.14-0.22\\
	& 6 &   0.07 - 0.11 \\
	 &  17  &   0.03-0.04  \\
	 & 25  &   0.02 - 0.03 \\
\hline
\hline
{\bf SMC} & 	& 	& 	\\
7.3  	& 0.3 &   0.24  &  Traditional Models \\
	& 0.5 &  	0.15	 & Max Dynamical Mass $< $3 kpc\\ 
	& 2   &  0.04     &  B12 \\ 
	& 4.2  & 0.02 	    &  Mean $\Lambda$CDM$^c$\\ 
\hline
17.3-40.3$^b$ &  0.3  & 0.58-1.34 &   \\
	& 0.5    &   0.35-0.81      &     \\ 
	& 3    &   0.06-0.13   &   \\
	& 4.2  & 0.04-0.1  &   \\
\noalign{\smallskip}\hline\noalign{\smallskip}
\end{tabular}
\newline
$^a$ Gas mass/ (Stellar Mass + Gas Mass)
$^b$ Including half the mass in the Magellanic Stream (total $2\times10^{9}$ M$\odot$) \citep{Fox14} at a distance of 55 kpc, or where
half the stream is at 100 kpc (total $6.6\times10^{9}$ M$_\odot$). 
$^c$ Mean value from relations in \citet{Mos13} for LMC/SMC stellar masses. 
\end{table}

\section{Plausible Orbital Histories for the LMC} 
\label{sec:4}

\begin{figure}[htbp]
\includegraphics[scale=.45, angle=90]{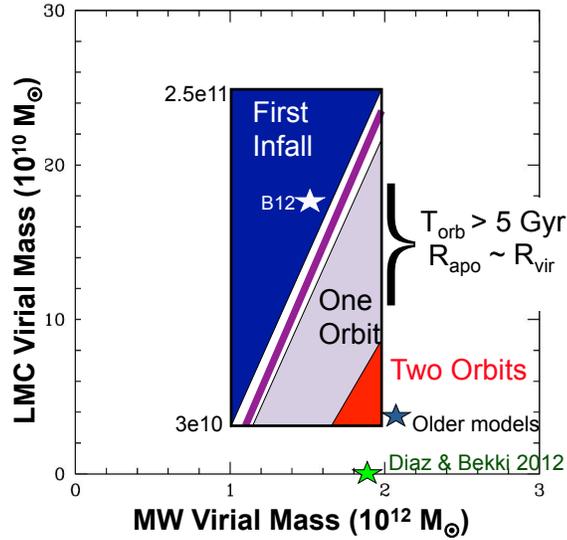}

\caption{Typical orbital histories for the LMC are indicated as a function of LMC and MW mass. 
Orbits are determined by searching the 4$\sigma$ proper motion error space in Monte Carlo
fashion and computing the mean number of pericentric passages completed within 10 Gyr.  
High mass LMC models experience greater dynamical friction and are consequently on 
more eccentric orbits, yielding first infall solutions (dark blue regions).
Lower mass models allow for orbits where the LMC has made one pericentric passage
(light purple regions); such solutions typically have orbital periods of order 5 Gyr. 
High MW mass and low LMC mass combinations are required for the LMC to have completed
more than one orbit. The mass of the LMC needs to be larger than $3\times 10^{10}$ M$_\odot$
in order to account for LMC stars located at distances of $\sim$15 kpc from the LMC center
and greater than $6\times 10^{10}$ M$_\odot$ for the baryon fraction of the LMC to approach 
$\sim$5\%.  Most studies have not considered such high mass models, apart from B12. 
  } 
\label{fig:1}     
\end{figure}

Following the methodology outlined in K13
and considerations for the mass of the MW and the mass and velocity of the LMC described in 
\ref{sec:3}, the orbit of the LMC can be integrated backwards in time. 
For various combinations of MW and LMC mass, Monte-Carlo 
drawings from the LMC's 4$\sigma$ velocity error distribution are 
used to explore plausible orbital histories.   
Figure~\ref{fig:1} shows the resulting mean orbital solutions for each MW/LMC mass combination. 
The LMC is considered to be on its first infall if it has not experienced more than
one pericentric approach within the past 10 Gyr.   In all cases the LMC has made at least one 
pericentric passage, since it is just at pericenter now.  

The illustrated dependence on LMC mass explains the discrepancy 
in the literature concerning different orbital solutions. Most studies have adopted low 
mass LMC models, which allows for orbital solutions with lower eccentricity.  
For example, \citep{Zha12} explore a variety of MW models to constrain the orbital history 
of the LMC, but they consider only one LMC mass model of $2 \times 10^{10}$ M$_\odot$. 
Numerical models of the Magellanic System by
 \citep{Diaz12, Diaz11} consider a total mass of only $10^{10}$ M$_\odot$, which is 
discrepant with the dynamical mass determined from the LMC's rotation curve.
 Other recent orbital studies adopt LMC masses of $3 \times 10^{10}$ M$_\odot$ 
\citep[e.g.,][]{Ruz10,Sha09}.  
These low masses are at odds with mass estimates from arguments 
about the baryon fraction of the LMC, which require a minimum 
dark matter mass of at least $6 \times 10^{10}$ M$_\odot$ ($\S$\ref{sec:3}).

The numerical models presented in \citep[][hereafter B10 and B12]{B10, B12} were designed 
to account for cosmological expectations, and thus adopt an LMC
mass of $1.8 \times 10^{12}$ M$_\odot$.

Note that while lower halo masses do allow for solutions where the LMC has completed 
one orbit about the MW, the mean orbital period  
is 5 Gyr. This timescale is much longer than the age of the Magellanic Stream. 
Furthermore, in this study, the MW mass is assumed to be static in time. If the mass 
evolution of the MW were included, the orbital period would be even longer.

\section{The LMC-SMC Binary} 
\label{sec:5} 

At a distance of $\sim$50 kpc from the Galactic center, the LMC and SMC are our closest example of an 
interacting pair of dwarf galaxies. Evidence of their ongoing interaction is clearly illustrated by the 
existence of the Magellanic Bridge that connects the two galaxies. This structure likely formed during their
last close approach $\sim 100-300$ Myr ago \citep[B12,][]{GN96}.  

The tidal field of the MW makes it statistically improbable that the LMC could have randomly captured the 
SMC some 300 Myr ago while in orbit about the MW.  It is more likely that the two galaxies were accreted 
as a binary; but the longevity of their binary status is unclear. 

All models for the Magellanic Stream and Bridge invoke tidal interactions between the MCs 
to some degree. The MCs must therefore have interacted 
for at least the lifetime of the Stream.  Based on the current high rate of ionization of the Stream \citep{Weiner96} and large
extended ionized component \citep{Fox14}, it is unlikely that the Stream could have survived as a neutral HI structure 
for more than 1-2 Gyr \citep{Bland07}.

The star formation histories (SFHs) of the MCs also suggest a common evolutionary history.
\citet{Weisz13} illustrate that $\sim$4 Gyr ago, the SFHs of both the LMC and SMC appear to increase in concert.
It is thus reasonable to assume the MCs have maintained a binary status for at least the 
past 4 Gyr. 
\\
\newline
\noindent {\bf Plausible Orbital Histories for the LMC-SMC Binary}

\begin{figure}[btbp]
\includegraphics[scale=.45,angle=90]{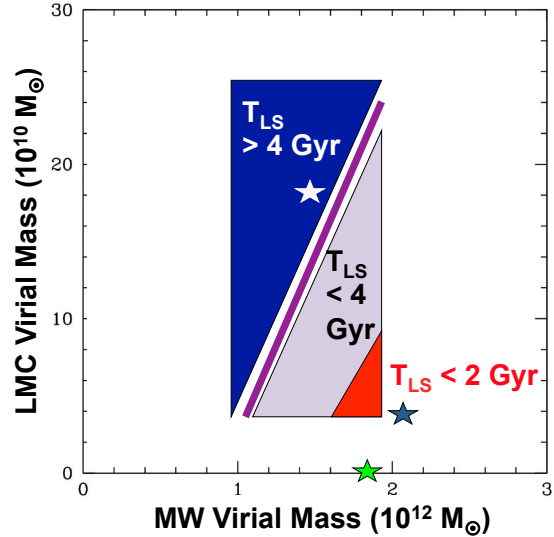}

\caption{Similar to Figure \ref{fig:1}, except now the SMC is also included in the orbital 
analysis; its 4$\sigma$ proper motion error space is also searched in Monte Carlo fashion. 
Colored regions represent the same LMC orbital solutions as in Figure\ref{fig:1}, 
but here the longevity of the LMC-SMC binary is indicated.  The SMC is assumed to be bound
to the LMC if the relative velocity between the Clouds is less than the local escape speed of the 
LMC at any point in the past.  Binary solutions are only viable in low/intermediate mass MW 
models + high LMC models; these are necessarily first infall orbits.    Dynamical friction between
the Clouds is not included in this analysis, but is expected to decrease the longevity of the 
binary orbit.  
  } 

\label{fig:2}       
\end{figure}

The orbital analysis presented in Figure \ref{fig:1} is revisited, exploring the same 
LMC/MW mass range, but this time also searching the 4$\sigma$ proper motion error space 
of the SMC in addition to that of the LMC.   The goal is to identify the combinations of LMC and 
MW mass that allow for the relative velocity of the MCs to be lower than the local escape 
speed of the LMC for some time in the past (not necessarily including today).  
Figure \ref{fig:2} illustrates the resulting mean longevity of the LMC-SMC binary as a function of LMC and 
MW mass. 

The action of MW tides are detrimental to the longevity of the Magellanic binary. 
As the mass of the MW increases, its tidal field is stronger and  
thus long-lived binary configurations are rare.  In particular, no long-lived solutions are found if the 
mass of the MW is greater than $2\times 10^{12}$ M$_\odot$.  This places an interesting upper 
bound on the virial mass of the MW. If the MW is $> 2\times 10^{12}$ M$_\odot$, the MCs could not have 
interacted for an appreciable amount of time in the past and their current proximity would be 
a random happenstance.  Few
upper bounds exist on the mass of the MW, making this a novel constraint. 

The SMC has historically been modeled in a circular orbit
about the LMC, with a relative velocity of order 60 km/s \citep{GN96}.  However, the new HST measurements reveal
that the relative velocity between the Clouds is significantly larger. At a relative velocity of 128 $\pm$ 32 km/s (K13),
the SMC is moving well above the escape speed of the LMC if its total mass is $3 \times 10^{10}$ M$_\odot$ 
(V$_{esc} \sim$110 km/s). This relative velocity measurement is a robust result that
has been confirmed by other teams \citep{Vieira10, Piatek08} 
and has not changed substantially from the earlier HST results \citep{K06b}. This high speed makes it very 
difficult to maintain a long-lived binary, unless the LMC is substantially more massive than traditionally modeled. 

The preferred configuration for a long-lived LMC-SMC 
binary is a low/intermediate mass MW + a high mass LMC.  This is exactly the opposite requirement for orbital solutions 
where the LMC makes at least one orbit about the MW.  In fact, the colors in Figure \ref{fig:2}
correspond to the same as those in Figure \ref{fig:1}; all binary solutions that last longer than 4 Gyr are first
infall solutions.   

The high relative velocity between the Clouds implies that the SMC is on an eccentric orbit about the 
LMC.  Such binary configurations are easily disrupted by MW tides, meaning that even one previous 
pericentric passage is sufficient to have destroyed the binary.  
This result is consistent with the fact that only 3.5\% of MW type galaxies host both an LMC and 
SMC stellar mass analog \citep{Liu11}.  Similarly, statistics from cosmological 
simulations find only 2.5\% MW type dark matter halos host both LMC and SMC mass analogs \citep{BK11}.
Our MW galaxy is thus an oddity in that it hosts two massive satellites in close proximity to each other.
However, this rare configuration can be understood if the MCs have only recently passed pericentric 
approach for the first time; only now are MW tides operating to disrupt this configuration. 

This study implies that all existing models in the literature that invoke the new HST proper 
motions in combination with low mass LMCs 
{\it do not allow for long-lived LMC-SMC binary solutions}.  In particular, because of the high 
speeds, no binary LMC-SMC solutions can exist in a MOND framework \citep{Zhao13}.

\section{Further Support for a First Infall of a Binary LMC/SMC }  

The simplest argument in favor of a recent infall is the unusually high gas fractions of both the LMC and SMC, given 
their proximity to the MW.  \citet{vdBergh06} conducted a morphological comparison of the satellites of 
the MW and M31, finding that the L/SMC are the only two gas-rich dwarf Irregulars
at close Galactocentric distance to their host. There are numerous environmental factors that work
to quench star formation and morphologically 
change galaxies after they become satellite galaxies of massive hosts (see chapter by Carraro).
It is thus remarkable that a satellite 
such as the SMC could have retained such a high gas content if it were accreted over 5 Gyr ago. 

In this volume, Burkert discusses the gas consumption timescale for galaxies; upon accretion 
satellites are cut off from their gas supply and thus their star formation rates should decline over time. 
However, Gallart (also in this volume) has illustrated that the star formation rates of the LMC 
have been {\it increasing}, with no signs of quenching over the past 4 Gyr until very recently. 

Taken together these arguments strongly support a scenario where the Clouds are on their first 
infall to our system, having only been within the virial radius of the MW for the past 1-2 Gyr.

\section{ Lessons from the Magellanic Clouds }

The Magellanic Clouds are recent interlopers in our neighborhood. 
This statement is a consequence of the dramatically improved proper motions of the MCs 
made using the HST by K13 and 
our evolving understanding of the structure of dark matter halos. Specifically, these factors have forced us to 
reconsider the total dark matter masses of the MCs. 
The baryon fractions of the MCs and imply that their total masses
must be at least a factor of 10 larger than traditionally modeled.  Dynamical friction then requires their 
orbits to be highly eccentric, preventing short period orbits.  Finally, the existence 
of a high relative velocity, LMC-SMC binary today strongly argues against their having completed a previous pericentric 
approach about our Galaxy, as MW tides can efficiently disrupt such tenuous configurations. 

A first infall solution implies that the MCs have interacted as a binary pair prior to accretion; in B10 
and B12 we argued that tidal interactions between the pair gave rise to the formation of the Magellanic 
Stream, Bridge and Leading Arm. As such,  interactions between dwarfs galaxies
 are important drivers of their evolution and may explain the existence of Magellanic 
Irregular type galaxies (i.e. LMC analogs) in the field. 

However, this scenario creates a large challenge for the theory that the satellite 
galaxies of the MW occupy a unique orbital plane and/or were formed in a common event (see 
the chapter in this volume by Pavel Kroupa).  To be more concise, the biggest challenge posed to 
this plane of satellites and MOND orbital histories constructed for the MCs is 
the current gas fraction of the SMC and the consequent requirement that the {\bf SMC must be on a first infall}.  
Orbits for the SMC in MOND have small periods and apocenters \citep{Zhao13}.  Such 
orbital solutions cannot explain the absence of quenching in the star formation history of the SMC 
and the fact that there is currently just as much gas in the SMC as there is in its much larger 
companion the LMC.

\bibliographystyle{mn2e}
\bibliography{BeslaSeychelles}

\end{document}